\documentclass[prl,twocolumn,showpacs,amsmath,amssymb]{revtex4}
\usepackage{graphicx}

\begin{document}

\title{$J/\psi$ and $\eta_c$ in  the Deconfined Plasma from Lattice QCD}
\author{M. Asakawa}
\email{yuki@ruby.scphys.kyoto-u.ac.jp}
\affiliation{Department of Physics, Kyoto University, Kyoto 606-8502, Japan}
\author{T. Hatsuda}
\email{hatsuda@phys.s.u-tokyo.ac.jp} 
\affiliation{Department of Physics, University of Tokyo,
Tokyo 113-0033, Japan}

\begin{abstract}
Analyzing correlation functions of charmonia
at finite temperature ($T$)  
on $32^3\times$(32$-$96) anisotropic lattices 
by the maximum entropy method (MEM),
we find that $J/\psi$ and $\eta_c$ survive
as distinct resonances in the plasma
even up to $T \simeq 1.6 T_c$ and that they eventually dissociate 
between $1.6 T_c$ and $1.9 T_c$ ($T_c$ is the 
critical temperature of deconfinement).
This suggests that the  deconfined plasma is non-perturbative
enough to hold heavy-quark bound states.
The importance of having sufficient number of
temporal data points in
MEM analyses
is also emphasized.
   
\end{abstract}
\maketitle

Whether hadrons survive even in the deconfined quark-gluon plasma
is one of the key questions in   
quantum chromodynamics (QCD). 
This problem was first examined 
in \cite{HK85} and \cite{DT85} in different contexts.
In the  former, it was   shown that collective
$q$-$\bar{q}$ excitations with
a low mass and a narrow width
in the $\pi$-$\sigma$ channels exist even 
above $T_c$ (the critical temperature) from
analyses of the spectral functions
in the Nambu-Jona-Lasinio model.
The fate of heavy mesons 
such as $J/\psi$ in the deconfined plasma  was also investigated in  
a phenomenological potential picture taking into account the Debye 
screening \cite{MS86}.
In general, there is no a priori reason to believe that
the dissociation of bound states 
should take place exactly at the phase transition point \cite{NSR85}.
    
Experimentally, measurements of
dileptons (diphotons) in heavy ion collisions 
may provide a clue to the properties of
vector (pseudo-scalar) mesons 
in hot/dense matter. Indeed, data from CERN SPS
indicate anomalies in
the dilepton spectra 
relevant to $\rho$ and $J/\psi$.  
Also, RHIC is going to produce ample
data of dileptons in a few years  \cite{QM02}.
 
From the theoretical point of view, the 
spectral function (SPF) at finite temperature $T$, which
has all the information of in-medium 
hadron properties, is a key quantity to be studied. 
Recently, the present authors have shown \cite{nah99} that 
the first-principle lattice QCD simulation
of SPFs is possible  by utilizing the maximum
entropy method (MEM). 
We have also formulated the basic concepts and applications of MEM
on the lattice  at $T=0$ and $T\neq 0$ in \cite{ahn01}.
The purpose of this Letter is to report our
latest results of SPFs of low lying charmonia
($J/\psi$ and $\eta_c$) above $T_c$ extracted from 
quenched QCD simulations on anisotropic lattices.
We find that $J/\psi$ and $\eta_c$  
survive even up to $T \simeq 1.6 T_c$ and 
disappear between 1.6$T_c$ and 1.9$T_c$.
This together with our previous results for  
$s\bar{s}$ mesons on the same lattice \cite{ahn02} suggests
that the system is still strongly interacting above $T_c$. 
   
To draw the above conclusion with a firm ground,
we put special emphases on
[I] the MEM error analysis of the resultant SPFs
and [II] the sensitivity of the SPFs to $N_{\rm data}$
(the number of the temporal data points 
adopted in MEM).
These tests are crucial to prevent fake generation
and/or smearing of the peaks and must be always carried out
as emphasized in \cite{ahn01,ahn02}. 
 
Let us first summarize the basic formulation of MEM 
applied to lattice data at finite $T$ \cite{ahn01}.  
We consider the Euclidean correlation function of  
the local interpolating operator 
$J_{l}(\tau,{\bf x}) = \bar{c}i\gamma_l c$
with $l=1,2,3$ for $J/\psi$ and $l=5$
for $\eta_c$.
The spectral decomposition of the correlator in the imaginary time
$0 < \tau < 1/T $ reads
\begin{eqnarray}
D (\tau) 
=  \int \langle J_l (\tau,{\bf x}) J_l ^{\dagger}(0) \rangle d^3x 
 = \int_{0}^{\infty} \!\! K(\tau,\omega) A(\omega) d\omega ,
\label{KA2}   
\end{eqnarray} 
where $\omega$ is a real frequency and
$A(\omega)$ is the spectral function. The sum over $l = 1,2,3$ is
taken for $J/\psi$. $K(\tau,\omega)$ is
the integral kernel, 
$K(\tau,\omega)=(e^{-\tau\omega}+e^{-(1/T-\tau)\omega})/(1-e^{-\omega/T})$.
For simplicity, we take the three momentum   of the 
correlation function to be zero.
  
Monte Carlo simulations provide $D(\tau_i)$ with statistical error
on a discrete set of temporal points $\tau_i $.
Although there exist infinitely many  $A(\omega)$ which give
the same $D(\tau_i)$ through the integral Eq.(\ref{KA2}),
MEM provides a way to select a {\em unique} $A(\omega )$ without
introducing parametrization of SPF and
to give its statistical significance on the basis of
the Bayes' theorem. The most probable
$A(\omega )$ given lattice data $D$ is obtained by
maximizing the conditional probability $P[A|D] \propto  e^{\alpha S - L}$,
where $L$ is the standard likelihood function and
$S$ is the Shannon-Jaynes entropy:
\begin{eqnarray}
S = \int_0^{\infty} \left [ A(\omega ) - m(\omega )
   - A(\omega)\log \left ( \frac{A(\omega)}{m(\omega )} \right ) \right ]
   d\omega . 
  \label{SJ-entropy}
\end{eqnarray}
Here $\alpha$ is a parameter
dictating the relative weight of $S$ and $L$.
The statistical significance (error) of the resultant $A(\omega )$
is estimated by the second variation, $(\delta/\delta A)^2 P[A|D]$.
The default model $m$ in Eq.(\ref{SJ-entropy}), the first estimate of which 
is obtained from the perturbative QCD calculation of
$A(\omega \gg 1 {\rm ~GeV})$,
may be chosen so that the MEM errors become minimum.    
The final result $A(\omega )$ is given by
the weighted average over $\alpha$: 
$A (\omega) = \int A_{\alpha}(\omega) \   P[\alpha|D m]  \ d\alpha $,
where $A_{\alpha}(\omega)$ is obtained by
minimizing $P[A|D] $ for
a fixed $\alpha$. 
The conditional probability $P[\alpha|Dm]$
can be calculated by using the Bayes' theorem and  
the lattice data. Thus $\alpha$ is eventually integrated out
and does not appear in the final result.
  
MEM has been successfully applied to the lattice QCD data at $T=0$
to extract the parameters of
the ground and excited state hadrons \cite{nah99,ssha02,Yamazaki02}.
On the other hand, applications to $T\neq 0$ system  have been known to be
a big challenge \cite{ahn01}, although there are some studies
at high $T$ \cite{kp02}.
The difficulty originates from the fact that
the temporal lattice size $L_{\tau}$ is restricted as
$L_{\tau}=1/T=N_{\tau} a_{\tau}$, where $a_\tau$ ($N_{\tau}$) is the
temporal lattice spacing
(the number of the temporal lattice sites). 
Because of this, it becomes more difficult to keep enough $N_{\rm data}$
to obtain reliable SPFs as $T$ increases.
In other words, simulations up to a few times $T_c$ with
$N_{\rm data}  $ as large as 30
\footnote{Our previous analysis on a $40^3\times30$ isotropic
lattice ($\beta = 6.47$ with  160 gauge configurations)
shows that $N_{\rm data}$ should be 28 or more to obtain
reliable SPFs \cite{ahn02}.
This number depends on lattice parameters and
statistics but gives a guide to
choose appropriate $N_{\tau}$ and $N_{\rm data}$ at finite $T$.
Therefore, data on lattices with small temporal grids such as
$N_{\tau} =12-16$ \cite{kp02} need to be examined carefully whether
they pass the test [II].} require fine lattices 
at least in the temporal direction 
(anisotropic lattices). 
 
On the basis of the above observation, we have
carried out  quenched simulations  with $\beta=7.0$
on $32^3\times N_\tau $  anisotropic lattice.
The renormalized anisotropy is
$\xi = a_{\sigma}/a_{\tau}=4.0$ with $a_{\sigma}$ being the
spatial lattice spacing.
We take the naive plaquette gauge action and the standard
Wilson quark action.
The corresponding bare anisotropy $\xi_0=3.5$ is 
determined from the data given in \cite{karsch_aniso,klassen}.
The fermion anisotropy
$\gamma_F \equiv \kappa_\tau /\kappa_\sigma$ with
$\kappa_\sigma$ ($\kappa_\tau$) being
the spatial (temporal) hopping parameter
is determined by comparing the temporal and
spatial effective masses of
the pseudoscalar and vector mesons 
on a $32^2 \times 48\times 128$ lattice. 
$a_\tau = a_{\sigma}/4 = 9.75\times 10^{-3}$ fm
is determined from the $\rho$ meson mass in the chiral limit
\footnote{The Polyakov-loop susceptibility
has a sharp peak around $N_\tau = 80 (72)$, which corresponds to
$T_c = 253 (281)$ MeV,
consistent with the known $T_c$ in pure gauge theory, $271 \pm 2$ MeV.}.
The physical lattice size in the spatial direction
$L_{\sigma} = 1.25 $ fm can well accommodate
$J/\psi$ or $\eta_c$, whose root mean square radius
is about 0.5 fm. $N_\tau $, 
the corresponding $T/T_c$, 
and the number of gauge configurations $N_{\rm gauge}$
are summarized in TABLE I. Calculations close to $T_c$
($T/T_c=1.04$ and $0.93$) are also currently undertaken.
Gauge configurations are generated by the pseudo
heatbath and over-relaxation algorithms with a ratio 1:4.
Initially, the gauge field is thermalized with 10000 sweeps
and, then each configuration is separated by 1000 sweeps.
 
\begin{table}[htbp]
 \caption{Temporal lattice size,
 the corresponding temperature,
 and the number of gauge configurations in our simulation.} 
 \begin{center}
  \begin{tabular}{|c|c|c|c|c|c|}
    \hline
   $N_{\tau}$ &   32   &  40    & 46     & 54    & 96   \\
    \hline  \hline
   $T/T_c$  &   2.33  & 1.87     & 1.62  & 1.38   &   0.78   \\
    \hline
   $N_{\rm gauge}$  &  141  & 181   & 182   & 150   &     194    \\
    \hline
  \end{tabular}
 \end{center}
\end{table}
 
We have measured the point-point $q\bar{q}$ correlation functions in the
scalar (S), pseudo-scalar (PS), vector (V), and axial-vector (AV)
channels for the spatial hopping parameters
$\kappa_\sigma = 0.08285, 0.0850, 0.0853,$ and 0.08545.
In this Letter, we focus our attention on
the heaviest quark mass ($\kappa_{\sigma} = 0.08285$
with $\gamma_F = 3.476$), and study
the V ($J/\psi$) and PS ($\eta_c$) channels. 
The masses determined on the $T=0$ lattice
($32^2 \times 48\times 128$) are
$m_{J/\psi}=3.10$ GeV and $m_{\eta_c}=3.03$ GeV.
  
In all the figures below, we adopt
the continuum kernel, $K(\tau,\omega)$, defined after Eq.(\ref{KA2}).
Use of the lattice kernel \cite{ahn01,kp02}
does not lead to any appreciable difference.
This is because our lattice spacing is quite small,
$a_{\tau} \sim$ 0.01 fm \footnote{
We discretize the integral Eq.(\ref{KA2}) into 1200 points
between $\omega=0$ and $\omega_{\rm max}$=30 GeV.
Results do not change for larger values of $\omega_{\rm max}$.}.
Following \cite{ahn01}, we define dimensionless
SPFs:  $A(\omega) = \omega^2 \rho(\omega)$ for $\eta_c$ and
$ A (\omega) = 3\omega^2 \rho(\omega)$ for $J/\psi$.
If the temporal distance between the source and the
sink is closer than $\xi a_{\tau}$,
lattice artifact due to anisotropy would appear
in the SPFs for $\omega \ge \pi/\xi a_\tau$.
To avoid this, we exclude
the six points near the edge ($\tau_i = 1, 2, 3$ and
$N_\tau- 3, N_\tau -2, N_\tau-1$) and adopt
the points $\tau_i= 4, 5, \cdots$ and $N_{\tau}-4, N_{\tau}-5, \cdots$
until we reach the total number of points $N_{\rm data} (<N_{\tau}-7)$.

The current operator on anisotropic lattice $J_l^{\rm LAT} (x)$
and that in the continuum  $J_l^{\rm CON} (x)$ are
related by
$J_l^{\rm LAT} = \sqrt{a_\tau a_\sigma^5}J_l^{\rm CON}
/(2Z_l \sqrt{\kappa_\tau \kappa_\sigma})$, where $Z_l$ is
a nonperturbative renormalization constant.
The default model $m(\omega)$ extracted from the
perturbative evaluation of $A(\omega \gg 1 {\rm ~GeV} )$
also depends on $Z_l$ \cite{ahn01}.
We vary $Z_l$ from 1 (the value in the weak coupling limit)
and the known value at  $\beta = 6.0$ on
the isotropic lattice 
\cite{ahn01,goeckeler99}.
The central values of $m$ thus obtained
are $m(\omega) /(3\omega^2) = 0.40$ and $m(\omega)/\omega^2 = 1.15$
for the V and PS channels, respectively.
We use these values throughout the following figures.
We have checked that our conclusions are not modified within the
variation of $Z_l$ mentioned above.

\begin{figure}[b]
\centering
\includegraphics[scale=0.35]{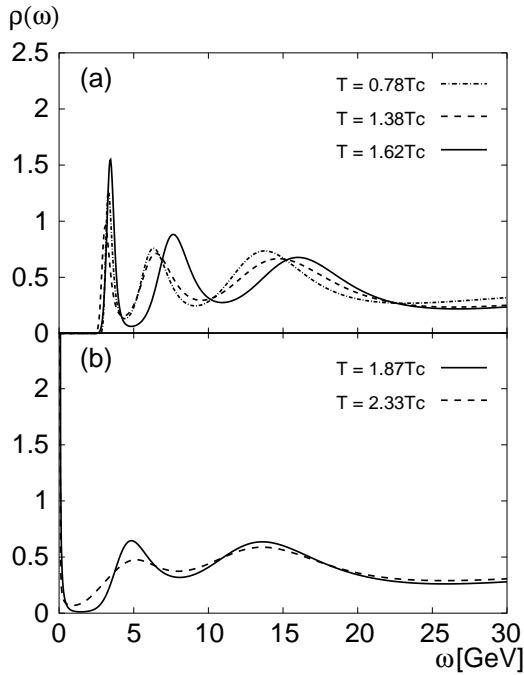}
\caption{Spectral functions for $J/\psi$ (a) for
 $T/T_c=0.78, 1.38$, and $1.62$
 (b) for $T/T_c=1.87$ and $2.33$.}
\label{p-np}
\end{figure}  
   
\begin{figure}[b]
\centering
\includegraphics[scale=0.35]{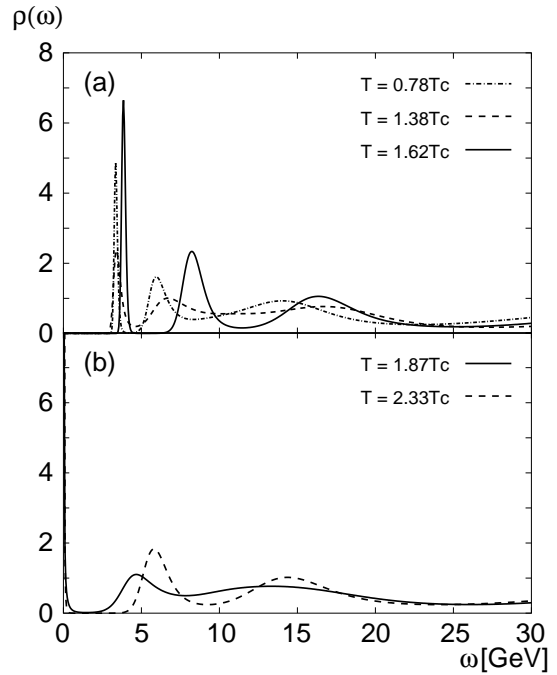}
\caption{Spectral functions for $\eta_c$
 (a) for $T/T_c=0.78, 1.38$, and $1.62$
 (b) for $T/T_c=1.87$ and $2.33$.}
\label{p-np-2}
\end{figure} 

Shown in Fig.\ref{p-np}  are $\rho(\omega)$s
for $J/\psi$ at $T/T_c = 0.78, 1.38$, and $1.62$ (Fig.\ref{p-np}(a))
and those at $T/T_c = 1.87$ and $2.33$ (Fig.\ref{p-np}(b))
(The corresponding $N_{\rm data}$  
used in these figures are $N_{\rm data}=89, ~40, ~34, ~33,$ and $25$
from low $T$ to high $T$).
If the deconfined plasma were composed of almost free
quarks and gluons, SPFs would show a smooth structure with
no pronounced peaks above the $q\bar{q}$ threshold.
To the contrary, we find a sharp peak near the zero temperature mass
even up to $T\simeq 1.6T_c$ as
shown in Fig.\ref{p-np}(a), while the peak 
disappears at $T\simeq 1.9 T_c$
as shown in Fig.\ref{p-np}(b).
The width of the first peak in Fig.\ref{p-np}(a) partly
reflects the unphysical broadening due to
the statistics of the lattice data and
partly reflects possible physical broadening
at finite $T$. At the moment, the former width of a few hundred MeV
seems to dominate and we are not able to draw
definite conclusions on the thermal mass shift and broadening.
  
The second and third peaks
in Fig.\ref{p-np}(a) may be related to the fermion doublers
as first pointed out for light mesons
at $T=0$ \cite{Yamazaki02}. This must be
checked by studying whether
the peak position scales as $1/a_{\sigma}$
by varying $a_{\sigma}$, which remains to be a future problem. 
Our error analysis shows that the peak around
$\omega = 0$ in Fig.\ref{p-np}(b)
is not significant in the present statistics.
The qualitative change 
of the spectral structure
between $T/T_c=1.62$ and 1.87, and other features seen in
the $J/\psi$ channel  are also observed in the $\eta_c$ channel
as shown in Fig.\ref{p-np-2}.
  
\begin{figure}[t]
\centering
\includegraphics[scale=0.35]{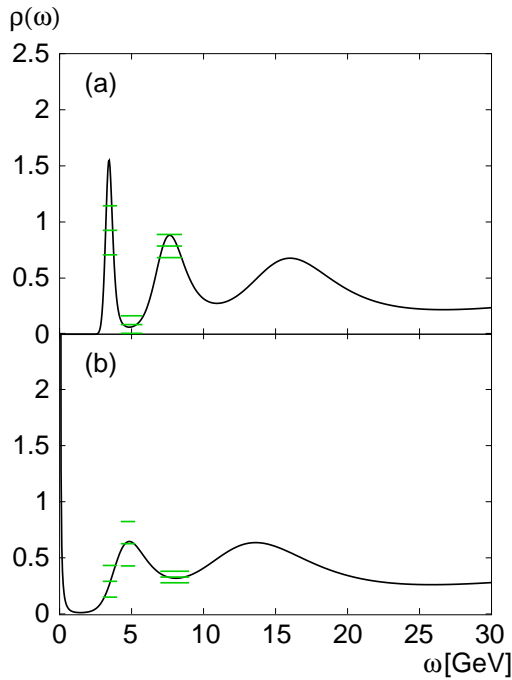}
\caption{ Spectral functions for $J/\psi$ with MEM errors
(a) for $T/T_c=1.62$
(b) $T/T_c=1.87$.}
\label{error}
\end{figure}
  
Let us now evaluate the reliability of the existence (absence) of
the sharp peak at $T/T_c = 1.62 (1.87)$
by the two tests [I] and [II] mentioned before.
The first test is 
the error analysis of the peak.
Shown in Fig.\ref{error} are the SPFs for $J/\psi$
at $T/T_c=1.62$ and 1.87 with MEM error bars
(The frequency interval over which
the SPF is averaged is characterized by the horizontal
position and extension of the bars, while the mean value and the 1$\sigma$
uncertainty of the integrated strength within the interval are
characterized by the heights of the bars).
The sharp peak at $T =1.62 T_c$ is statistically significant, and
the absence of the peak at the same position
at $T =1.87 T_c$ is also statistically significant.
The same features are also observed for $\eta_c$.
   
\begin{figure}[t]
\centering
\includegraphics[scale=0.35]{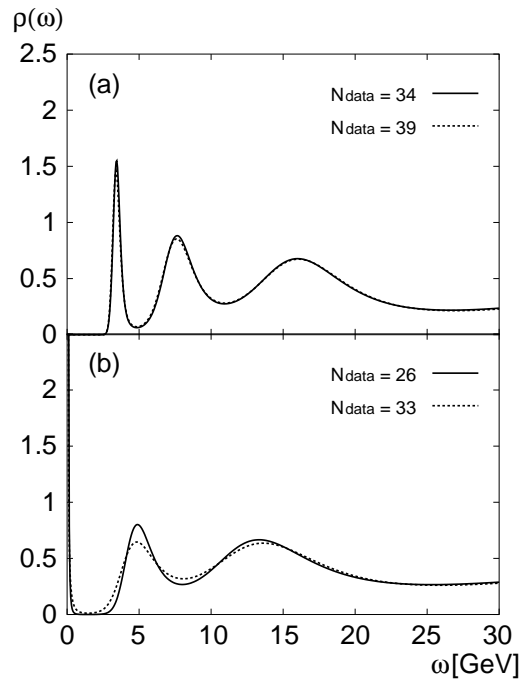}
\caption{Comparison of the SPF for $J/\psi$
(a) for $ N_{\rm data} = 34$ and 39
  with $N_{\tau}=46$ $(T/T_c=1.62)$
(b) for $N_{\rm data} =26 $ and 33
  with $N_{\tau}=40$ $(T/T_c=1.87)$. }
\label{N-dep}
\end{figure}

The second test is the $N_{\rm data}$ dependence of the SPFs.
Shown in Fig.\ref{N-dep}(a) is
a comparison of the SPF 
obtained with $N_{\rm data}=34$ and
that with 
$N_{\rm data}= 39$ for the same temperature
$T =1.62 T_c$ $(N_{\tau} =46)$.
The two curves  are almost identical with each other
(Note that the maximum number of temporal data
available in this case is $46-7=39$).
Fig.\ref{N-dep}(b) shows SPFs obtained with
$N_{\rm data}=26$ and 33 for
a higher temperature 
$T =1.87 T_c$ $(N_{\tau}=40)$. The maximum number
of data points available is $40-7=33$ in this case.
Again the two curves are almost identical.
Therefore, the qualitative change of the SPF between
$T=1.62 T_c$ and 1.87$T_c$ is the real thermal effect  
and not caused by the artifact of the
insufficient number of data points.
    
Here we make a brief comment on related works
aiming at studying the charmonia above $T_c$
using MEM \cite{Umeda02,BI02}. In contrast to our
large temporal grids
($N_{\tau}=46$ for $T=1.62T_c$),
$N_{\tau}$ in these papers are  3-4 times smaller
($N_{\tau}= 17$ in  \cite{Umeda02}
and $N_{\tau}= 12$ in \cite{BI02} at $T =1.62 T_c$).
For such small $N_{\tau}$ with point-point correlation,
the reliability tests [I] and [II] need to be done
before drawing  physics conclusions.
There is indeed evidence 
that $N_{\tau}= 17$ fails the test [II] as
shown in \cite{Umeda02}.
     
In summary, we have extracted
the spectral functions of
$J/\psi$ and $\eta_c$ in the deconfined plasma
using lattice Monte Carlo data and the maximum entropy method.
In the quenched approximation,
the number of temporal sites $N_{\tau}$ is taken
as large as 46 and 40 for $T/T_c = 1.62$ and $1.87$, respectively.
Careful analyses of the MEM errors and the $N_{\rm data}$ dependence
of the results are carried out.
It is found that there are distinct resonances
up to $T \simeq 1.6T_c$ and they disappear between 1.6$T_c$
and  $1.9 T_c$.
This together with our previous results
on $s\bar{s}$ mesons \cite{ahn02} on the same lattice
indicates that the quark-gluon plasma is still strongly interacting
above $T_c$ so that it can develop hadronic resonances.
Whether these resonances found in quenched
simulation survive in full simulation with dynamical quarks
is an interesting future problem.
It is also important to unravel the nature of such resonances
by studying their spatial structure on the lattice \cite{spatial-corr}
or by using non-relativistic potential approaches \cite{POT-app}.
   
\begin{acknowledgments}
M.A. (T.H.) is partially
supported by the Grants-in-Aid of the Japanese Ministry of Education,
Science and Culture, No.~14540255 (No.~15540254).
Lattice calculations have been performed with the CP-PACS computer
under the ``Large-scale Numerical Simulation Program" of
Center for Computational Physics, University of Tsukuba.
We are particularly grateful to the encouragement and technical
support in using the CP-PACS computer from S. Aoki, K. Kanaya,
M. Tomida, A. Ukawa, and T. Yoshi\'e.

\end{acknowledgments}
 

\end{document}